\documentstyle[12pt]{article}

\newcommand{\EQ}{\begin{equation}}
\newcommand{\EN}{\end{equation}}
\begin{document}
\topmargin 0pt
\oddsidemargin=-0.4truecm
\evensidemargin=-0.4truecm
\renewcommand{\thefootnote}{\fnsymbol{footnote}}
\newpage
\setcounter{page}{1}
\begin{titlepage}
\begin{flushright}
IC/92/79\\
SISSA-83/92/EP\\
LMU-04/92\\
May 1992
\end{flushright}
\vspace*{-0.2cm}
\begin{center}
{\large PLANCK-SCALE PHYSICS AND NEUTRINO MASSES}
\vspace{0.5cm}

{\large Eugeni Kh. Akhmedov${}^{(a,b,c)}$
\footnote{E-mail: akhmedov@tsmi19.sissa.it, ~akhm@jbivn.kiae.su},
Zurab G. Berezhiani${}^{(d,e)}$
\footnote{Alexander von Humboldt Fellow}
\footnote{E-mail: zurab@hep.physik.uni-muenchen.de, ~vaxfe::berezhiani},\\
Goran Senjanovi\'{c}${}^{(a)}$}
\footnote{E-mail: goran@itsictp.bitnet, ~vxicp1::gorans}
\vspace*{0.4cm}

${}^{(a)}$\em{International Centre for Theoretical Physics,
I-34100 Trieste, Italy\\
${}^{(b)}$Scuola Internazionale Superiore di Studi Avanzati,
I-34014 Trieste, Italy\\
${}^{(c)}$Kurchatov Institute of Atomic Energy, Moscow 123182, Russia\\
${}^{(d)}$Sektion Physik der Universit\"{a}t M\"{u}nchen, D-8000 Munich-2,
Germany\\
${}^{(e)}$Institute of Physics, Georgian Academy of Sciences, Tbilisi
380077, Georgia\\}
\end{center}
\vglue 0.8truecm
\begin{abstract}
We discuss gravitationally induced masses and mass splittings of Majorana,
Zeldo-vich-Konopinski-Mahmoud and Dirac neutrinos. Among other implications,
these effects can provide a solution of the solar neutrino puzzle.
In particular, we show how this may work in the 17 keV neutrino picture.
\end{abstract}
\vspace{1.cm}
\end{titlepage}
\renewcommand{\thefootnote}{\arabic{footnote}}
\setcounter{footnote}{0}
\newpage
{\em A. Introduction}

It is commonly accepted, although not proven, that the
higher dimensional operators induced through the quantum gravity effects
are likely not to respect global symmetries. This is, at least in part,
a product of one's experience with black holes and wormholes. If so, it
becomes important to study the impact of such effects on various global
symmetries of physical interest. Recently, an attention has been drawn
to the issues of Peccei-Quinn symmetry \cite{Kolb1} and
global non-abelian symmetry relevant for the textures
\cite{Kolb2}. Here, instead, we study the possible impact of
gravity on the
breaking of lepton flavor and lepton number, more precisely its impact
on neutrino (Majorana) masses. It is clear that all such effects, being
cut off by the Planck scale, are very small, but on the other hand even
small neutrino mass can be of profound cosmological and astrophysical
interest.

We start with a brief review of a situation in the standard model where
such effects can induce, as pointed out by Barbieri, Ellis and Gaillard
\cite{BEG} (in the language of $SU(5)$ GUT), a large enough neutrino
masses to explain the solar neutrino puzzle (SNP) through the vacuum
neutrino oscillations. We also derive the resulting neutrino mass
spectrum and the mixing pattern, which have important implications
for the nature of the solar neutrino oscillations. From there on we
center our discussion on the impact of the mass splits between the
components of Dirac and Zeldovich-Konopinski-Mahmoud (ZKM) \cite{ZKM}
neutrinos. Our main motivation is the issue of
SNP, but we will discuss the cosmological implications as well. The
important point resulting from our work is a possibility to incorporate
the solution of the SNP in the 17 keV neutrino picture in a simple and
natural manner. Finally, we make some remarks on the see-saw mechanism
and also mirror fermions in this context.

\vspace{0.4cm}
{\em B. Neutrino masses in the standard model}

Suppose for a moment that no right-handed neutrinos exist, i.e. the
neutrinos are only in left-handed doublets. Barring accident cancellations
(or some higher symmetry), the lowest-order neutrino mass effective
operators are expected to be of dimension five:
\EQ
\alpha_{ij}\,l_{i}^{T}C\tau_{2}\vec{\tau}l_{j}\,
\frac{H^{T}\tau_{2}\vec{\tau}H}{M_{\rm{Pl}}}
\EN
where $l_{i}=(\nu_{iL}~ e_{iL})^{T}$, $H$ is the usual $SU(2)_{L}\times
U(1)$ Higgs doublet, $M_{\rm{Pl}}$ is the Planck mass $\approx 10^{19}$
GeV and $\alpha_{ij}$ are unknown dimensionless constants. The operator
(1) was first written down by Barbieri et al. \cite{BEG} who, as we have
mentioned before, based their discussion on $SU(5)$ GUT although strictly
speaking it only involves the particle states of the standard model.
If gravity truly breaks the lepton number and induces terms in (1),
neutrino may be massive even in the minimal standard model. This important
result of ref. \cite{BEG} seems not to be sufficiently appreciated in the
literature. As was estimated in \cite{BEG}, for $\alpha_{ij}\sim 1$ one
gets for the neutrino masses $m_{\nu}\sim 10^{-5}$ eV, which is exactly of
the required order of magnitude for the solution of the SNP through the
vacuum neutrino oscillations.

We would like to add the following comment here. The universality of
gravitational interactions makes it very plausible that all the
$\alpha_{ij}$ constants in (1) should be equal to each other: $\alpha_{ij}=
\alpha_{0}$. If so, the neutrino mass matrix must take the "democratic"
form with all its matrix elements being equal to $m_{0}\sim \alpha_{0}\cdot
10^{-5}$ eV. For three neutrino generations this pattern implies two
massless neutrinos and a massive neutrino with $m_{\nu}=3m_{0}$. The
survival probability of $\nu_{e}$ due to the $\nu_{e}\rightarrow \nu_{\mu},
\nu_{\tau}$ oscillations is
\EQ
P(\nu_{e}\rightarrow \nu_{e};t)=1-\frac{8}{9}\sin^{2}\left(
\frac{m_{\nu}^{2}}{4E}t \right )
\EN
For $m_{\nu}\sim 10^{-5}$ eV, the oscillations length
$l=4\pi E/m_{\nu}^{2}$ for the neutrinos with the energy $E\sim 10$ MeV is
of the order of the distance between the sun and the earth. Since eq. (2)
describes the large-amplitude oscillations, one can in principle get a
strong suppression of the solar neutrino flux. This, so-called "just so",
oscillation scenario leads to well defined and testable consequences
\cite{JS}.

We should add that if the relevant cut-off scale in (1) would be one or two
orders of magnitude smaller than $M_{\rm{Pl}}$ (as can happen in the string
theory, where the relevant scale may be the compactification scale),
$m_{\nu}$ could be as large as $10^{-4}-10^{-3}$ eV. One can distinguish two
cases then. For $10^{-10}<m_{\nu}^{2}<10^{-8}$ eV${}^{2}$ we are faced with
the conventional vacuum oscillations for which eq. (2) gives the averaged
$\nu_{e}$ survival probability ${\bar P}(\nu_{e}\rightarrow \nu_{e})
\simeq 5/9$.
For $m_{\nu}^{2}>10^{-8}$ eV${}^{2}$ the MSW effect \cite{MSW} comes into
the game. Although all three neutrino flavors are involved, one can readily
make sure that the resonant oscillation pattern can be reduced to an
effective two-flavor one. If the adiabaticity condition is satisfied, the
neutrino emerging from the sun is the massive eigenstate
$\nu_{3}=(\nu_{e}+\nu_{\mu}+\nu_{\tau})/\sqrt{3}$. Thus,
$P(\nu_{e}\rightarrow \nu_{e})=1/3$ in this case.

\vspace{0.4cm}
{\em C. ZKM and Dirac neutrinos}

The major point of the result (1) is that the emerging neutrino masses
are on the borderline of the range needed for the solution of the SNP.
For $\alpha_{ij}\ll 1$, the mechanism would not work. The situation changes
drastically if there is an additional mechanism of generating neutrino
mass. This is particularly interesting in the case of ZKM or Dirac
neutrinos. By ZKM we generically denote any situation with degenerate
active neutrino flavors $\nu_{iL}$ and $\nu_{jL}$ when the lepton charge
$L_{i}-L_{j}$ is conserved. In other words, the resulting state is a
four-component neutrino $\nu_{\rm{ZKM}}=\nu_{iL}+(\nu_{jL})^{c}$.
In the conventional Dirac picture one has
$\nu_{\rm{D}}=\nu_{L}+n_{R}$,
where $n_{R}$ is sterile and the conserved charge is just the particular
lepton flavor defined through $\nu_{L}$.

In any case, if one desires to have the oscillations between the components
of $\nu_{\rm{ZKM}}$ ($\nu_{\rm{D}}$), one has to break the
degeneracy, i.e. induce Majorana masses which violate the conserved charge
in question. This, as before, can be achieved through the gravity induced
operators of (1) and also in the same manner by
\EQ
(n_{R}^{T}Cn_{R})\left[\alpha_{n}\frac{H^{\dagger} H}{M_{\rm{Pl}}}+
\beta_{n}\frac{S^{2}}{M_{\rm{Pl}}}\right ],
\EN
where $S$ is any $SU(2)_{L}\times U(1)$ singlet scalar field, which may
or may not be present. Here, as throughout our analysis, we assume no
direct right-handed neutrino mass such as $n_{R}^{T}Cn_{R}\,S$ which is
implicit in our assumption of having a Dirac state
\footnote{The $n_{R}^{T}Cn_{R}\,S$ terms could be forbidden by a global
charge.}.

Let us focus first on the ZKM case. In general, the terms in (1) will
induce the split $\Delta m\leq 10^{-5}$ eV {}
\footnote{A ZKM neutrino with a small energy split between its
components is usually called pseudo-Dirac neutrino \cite{PD}.}.
The oscillation
probability depends on $\Delta m^{2}\sim m\Delta m$. For this
to be relevant for the SNP, one of the components $\nu_{i}$ or $\nu_{j}$
must be $\nu_{e}$ and therefore $m\leq 10$ eV \cite{LA}, or $\Delta
m^{2}\leq 10^{-4}$ eV${}^{2}$. Clearly, for any value of $m\geq 10^{-5}$
eV, this can provide a solution to the SNP through the vacuum
oscillations
\footnote{Although $\Delta m^{2}$ could easily be in the range $\Delta
m^{2}\simeq 10^{-8}-10^{-4}$ eV${}^{2}$, the MSW effect is irrelevant
for the SNP in this case since the vacuum mixing angle is practically
equal to $45^{\circ}$.}.

We would like to mention that $\Delta m^{2}$ in the above range can
be also relevant for another possible explanation of the SNP, namely,
resonant spin-flavor precession of neutrinos due to the transition
magnetic moment between $\nu_{i}$ and $\nu_{j}$ \cite{ALM}.

The same qualitative analysis holds true for the Dirac neutrino, the only
difference being the additional contribution of (3) to the Majorana
masses. The possible presence of the $S^{2}$ term (if $<S>\not=0$)
could modify drastically the predictions for $\Delta m$. Strictly speaking,
in a general case no statement is possible at all since $<S>$ could be
in principle as large as  $M_{\rm{Pl}}$. Of course, in the most
conservative scenario of no new Higgs fields above the weak scale, the
analysis gives the same result as for the ZKM situation. Oscillations
(or resonant spin-flavor precession) between the components of a Dirac
neutrino can also provide a solution to the SNP; however, the experimental
consequences for experiments such as SNO or Borex will be different.
Namely, the detection rates in the neutral current mediated reactions
will be reduced since the resulting neutrino is sterile.

Another important consequence of the induced mass splitting between the
components of a Dirac neutrino is a possibility to have a sterile neutrino
brought into the equilibrium through the neutrino oscillations at the time
of nucleosynthesis \cite{BD}. This has been analysed at length in
ref. \cite{ABST}, and can be used to place limits on $\Delta m_{ij}^{2}$
and neutrino mixings.
\newpage
%
{\em D. 17 keV neutrino}

A particularly interesting application of the above effects finds its
place in the problem of 17 keV neutrino \cite{PRO}. Although the very
existence of this neutrino is not yet established, it is tempting
and theoretically challenging to incorporate such a particle into our
understanding of neutrino physics.

Many theoretical scenarios on the subject were proposed; however,
it is only recently that the profound issue of the SNP in this picture
has been addressed. The problem is that the conventional scenario of three
neutrino flavors $\nu_{eL}$, $\nu_{\mu L}$ and $\nu_{\tau L}$ cannot
reconcile laboratory constraints with the solar neutrino deficit. Namely,
the combined restriction from the neutrinoless double beta decay and
$\nu_{e}\leftrightarrow\nu_{\mu}$ oscillations
leads to a conserved (or at most very
weakly broken) generalization of the ZKM symmetry: $L_{e}-L_{\mu}+L_{\tau}$
\cite{V}
\footnote{This lepton charge was first introduced in another context
in ref. \cite{P}.}. This in turn implies the 17 keV neutrino mainly
to consist of $\nu_{\tau}$ and $(\nu_{\mu})^{c}$, mixed with the Simpson
angle $\theta_{\rm{S}}\sim 0.1$ with the massless $\nu_{e}$. Clearly, in
this picture there is no room for the solution of the SNP due to neutrino
properties.

It is well known by now that the LEP limit on $Z^{0}$ decay
width excludes the existence of yet another light active neutrino.
However, the same in general is not true for a sterile neutrino
$n$. Of course, once introduced, $n$ (instead of $\nu_{\mu}$) can combine
with, say, $\nu_{\tau}$ to form $\nu_{17}$ or just provide a missing light
partner to $\nu_{e}$ needed for the neutrino-oscillations solution to the
SNP. The latter possibility has been recently addressed by the authors of
ref. \cite{X}.

The introduction of a new sterile state $n$ allows for a variety of
generalizations of a conserved lepton charge $L_{e}-L_{\mu}+L_{\tau}$.
This will be analyzed in detail in the forthcoming publication
\cite{ABST}. Here we concentrate on the simplest extension
$\hat{L}=L_{e}-L_{\mu}+L_{\tau}-L_{n^{c}}$
\footnote{The phenomenology of the system with the conserved
lepton charge $\hat{L}$ (with $n$ being an active neutrino of the fourth
generation) was analysed in \cite{L}.}
and assume the following physical states:
\EQ
\nu_{17}\simeq \nu_{\tau}+(\nu_{\mu})^{c},~~~\nu_{light}\simeq \nu_{e}+n
\EN
mixed through $\theta_{\rm{S}}$. In the limit of the conserved charge
$\hat{L}$, the light state is a Dirac particle and no oscillations are
possible which would be relevant for the SNP. Furthermore, the only allowed
oscillations are $\nu_{e}\leftrightarrow \nu_{\tau}$ and
$\nu_{\mu}\leftrightarrow n$ with $\Delta m^{2}\simeq (17$ keV$)^{2}$.
The situation changes drastically even with a tiny breaking of $\hat{L}$
and we show here how the potential gravitational effects in (1) and (3)
may naturally allow for the solution of the SNP without any additional
assumptions.

Clearly, the main impact of the above effects is to induce the mass
splittings between the $\nu_{\tau}$ and $\nu_{\mu}$ on one hand, and
$\nu_{e}$ and $n$ on the other hand. Recall that we expect these
contributions to be of the order of $10^{-5}$ eV or so, if no new scale
below $M_{W}$ is introduced. This tells us that
\EQ
\Delta m_{\nu_{\tau} \nu_{\mu}}^{2}\leq 10^{-1}~\rm{eV}^{2}, ~~~
\Delta m_{\nu_{e} n}^{2}\leq 10^{-4}~\rm{eV}^{2}
\EN
As we can see, the scenario naturally allows for the solution
of the SNP due to the vacuum oscillations ${}^{3}$
and furthermore predicts the $\nu_{\mu}\leftrightarrow \nu_{\tau}$
oscillations potentially observable in the near future. Notice that
although we discussed the case of only one sterile neutrino, any number
would do. Also, we should emphasize that the result is completely model
independent, as long as one deals with (almost) conserved charge $\hat{L}$.
However, a simple model can easily be constructed and will be presented
elsewhere \cite{ABST}.

\vspace{0.4cm}
{\em E. Discussion}

We have seen how gravitation may play a major role in providing neutrino
masses and mass splittings relevant for the SNP. Purely on dimensional
grounds, at least in the case of only left-handed neutrinos, the
Planck-scale physics induced masses and splittings are $\leq 10^{-5}$ eV.
The smaller they are, the larger the neutrino masses generated by some
other mechanism should be, in order to obtain large enough $\Delta m^{2}$.
For this reason a ZKM neutrino is rather interesting, especially with
cosmologically relevant $m_{\nu_{e}}$ close to its experimental upper
limit $\sim 10$ eV.

The situation is less clear in the case of a Dirac neutrino, since the
gravitationally induced mass term $m_{n}\,n^{T}Cn$ could in principle
give $m_{n}$ as large as $M_{\rm{Pl}}$. In fact, in this case it is hard
to decide whether one is dealing with a Dirac neutrino or actually with a
see-saw phenomenon \cite{SS}. The situation depends on the unknown aspects
at very high energies, i.e. whether or not the scale of the $n$ physics is
much above $M_{W}$.

The see-saw effects may be even more important if one is willing to promote
the whole $SU(2)_{L}\times SU(2)_{R}\times U(1)$ electroweak symmetry or,
in other words, if one is studying a parity conserving theory. This is a
natural issue in many GUTs, such as $SO(10)$ or $E_{6}$. Normally, in order
for the see-saw mechanism to work, one introduces a Higgs field which gives
directly a mass to $n$. In the spirit of our discussion it is clear that
gravitation may also do the job and, if so, one would expect $m_{n}\sim
M_{R}^{2}/M_{\rm{Pl}}$ where $M_{R}$ is the scale of the $SU(2)_{R}$
breaking, i.e. the scale of parity restoration. In other words, even with
large $M_{R}$, $m_{n}$ could be quite small allowing far more freedom
for light neutrino masses. Of course, the details are model-dependent
(i.e. $M_{R}$ scale dependent) and we do not pursue them here.

As we have seen in section {\em D}, all the cases relevant to the 17 keV
neutrino lead to the large mixing angle solution of the SNP. Before
concluding this paper, we would like to offer some brief remarks regarding
an interesting possibility of mirror fermions picture providing the desirable
MSW solution of this problem.

Imagine a world which mimics ours completely in a sense that these
mirror states have their own, independent weak interactions. In other words,
let gravitation be the only bridge between the leptonic sectors of the two
worlds. One would then have the following new mass operators in addition to
those in eq. (1):
\EQ
\alpha_{0}\,l_{Mi}^{T}C\tau_{2}\vec{\tau}l_{Mj}\,
\frac{H_{M}^{T}\tau_{2}\vec{\tau}H_{M}}{M_{\rm{Pl}}},~~~~
\alpha_{0}\,{\bar l}_{Mi}l_{j}\,
\frac{H^{\dagger}H_{M}}{M_{\rm{Pl}}}
\EN
where $M$ stands for mirror particles. Therefore, in addition to the 17 keV
neutrino described before there should be an analogous $(\nu_{17})_{M}$
neutrino in the mirror world, and another massless state much as like as the
massless state in the standard generalized ZKM picture. It turns out that,
due to the mixing in eq. (6), one of these two massless states picks up a
mass $\sim \alpha_{0}M_{W}^{2}M_{{\rm Pl}}^{-1}\theta^{-2}$ with
$\theta \simeq <H>$/$<H_{M}>$ whereas the
other still remains massless. Therefore they can oscillate into each
other with the mixing angle $\theta$, and so in
principle this can provide a solution of the SNP through the appealing MSW
effect. Namely, for $<H_{M}>\simeq 1-10$ TeV the mass difference $\Delta m^{2}$
could  easily be in the required MSW range. The above range of $<H_{M}>$
makes this scenario in principle accessible to the SSC physics.

\vspace{0.4cm}
{\em Acknowledgements}

We would like to thank I. Antoniadis, J. Harvey and G. Raffelt for
discussions and M. Lusignoli for bringing ref. \cite{L} to our attention.
E.A. is grateful to SISSA for its kind hospitality during the initial stage
of this work.\\

\noindent
{\em Note added}. After this work was completed we received a paper by
Grasso, Lusignoli and Roncadelli \cite{GLR} who also discuss gravitationally
induced effects in the 17 keV neutrino picture.

\end{document}